# Quantifying polarization across political groups on key policy issues using sentiment analysis

Dennies Bor[1], Benjamin Seiyon Lee[2], Edward J. Oughton[1,3] *

[1]College of Science, George Mason University, Fairfax, VA, USA.
[2]Department of Statistics, The George Mason University, Fairfax, Virginia, USA.
[3]Environmental Change Institute, University of Oxford, Oxford, Oxfordshire, UK.

*Corresponding author: Edward J. Oughton (e-mail: eoughton@gmu.edu)
Address: College of Science, George Mason University, 4400 University Drive, Fairfax, VA.

## Abstract

There is growing concern that over the past decade, industrialized democratic nations are becoming increasingly politically polarized. Indeed, elections in the US, UK, France, and Germany have all seen tightly won races, with notable examples including the 2016 Trump vs. Clinton presidential election and the UK's Brexit referendum. However, while there has been much qualitative discussion of polarization on key issues, there are few examples of formal quantitative assessments examining this topic. Therefore, in this paper, we undertake a statistical evaluation of political polarization for representatives elected to the US congress on key policy issues between 2021-2022. The method is based on applying sentiment analysis to Twitter data and developing quantitative analysis for six political groupings defined based on voting records. Two sets of policy groups are explored, including geopolitical policies (e.g., Ukraine-Russia, China, Taiwan, etc.) and domestic policies (e.g., abortion, climate change, LGBTQ, immigration, etc.). We find that out of the twelve policies explored here, gun control was the most politically polarizing, with significant polarization results found for all groups (four of which were P < 0.001). The next most polarizing issues include immigration and border control, fossil fuels, and Ukraine-Russia. Interestingly, the least polarized policy topics were Taiwan, LGBTQ, and the Chinese Communist Party, potentially demonstrating the highest degree of bipartisanship on these issues. The results can be used to guide future policy making, by helping to identify areas of common ground across political groups.

# 1. Introduction

Industrialized democratic countries have experienced a wave of political polarization over the past decade [1]–[3]. For example, a notable case includes the 2016 decision for the United Kingdom to leave the European Union by a relatively narrow margin (52% voting leave, and 48% voting remain) [4]. Moreover, a few months later, the 2016 United States Presidential Election saw a Republican victory in the electoral college with 46.1% of the vote, while the losing Democratic party did so while winning the popular vote with 48.2% [5]. There are similar contexts across a range of other advanced nations, including France, Italy, Spain, and numerous others, suggesting many societies are becoming increasingly ideologically divided [6]–[8].

Sadly, this division has spilled over into the development of new issues between different social groups, which is often popularized and referred to as the ongoing 'Culture War' [9]. This has most notably arisen in the United States between social movements on the extremes of the political spectrum [10]–[12]. These range from the Left to the Right, for example, from the reinvigoration of the century-old Anti-Fascist movement (Antifa) to the emergence of a new Alternative Right (Alt-Right) [10], [13], [14]. This polarization has been affiliated with a hollowing out of the political center ground, which has unfortunately spurred greater division and intolerance. An unfortunate consequence is that the free speech debate has been in retreat in many places. Moreover, intra-party acceptance of working on bipartisan activities has decreased [15]–[18].

Given that strong divisions can lead to more negative societal outcomes, we need to explore a variety of important questions in greater detail. These questions generally pertain to diagnosing polarization on specific current affairs issues. This insight can then be used to identify pluralist strategies which try to encourage bipartisan activities. Consequently, in this paper, we develop methods to answer the following research question:

1. How do the political narratives of different elected representatives compare on key current affairs issues, and to what degree are current political groupings polarized?

One population data source for gaining insight into political narratives is the social network site Twitter [19]. Increasingly, scientists are utilizing Twitter data and natural language processing (NLP) techniques to study public discourse on a wide variety of topics, including the economy, immigration, health, gun controls, clean energy, Covid-19 and vaccines, foreign policy, and climate change. Indeed, fundamental NLP techniques, such as sentiment analysis and topic modeling, can estimate the meaning of the unstructured text to provide a new systematic understanding of large quantities of data [20]–[26].

In Section 2 of this paper, a literature review is undertaken to evaluate existing political polarization studies, as well as the use of Twitter data and NLP in sentiment analysis. In Section 3, a method is subsequently articulated, enabling the research question to be investigated before the key results are reported in Section 4. Finally, a discussion is undertaken in Section 5, where the results are assessed in relation to the research question before conclusions are provided in Section 6.

## 2. Literature Review

The literature review is split into three sections. First, we evaluate the relevant literature which uses Twitter data to understand political polarization in Section 2.1. Secondly, the use of Twitter data in NLP studies is then assessed in Section 2.2. Finally, techniques available for conducting sentiment analysis, topic modeling, and other NLP methods for text data are reviewed in Section 2.3.

### 2.1 Review of literature on political polarization

In a democracy, where political division is a central tenant, the system functions by parties tolerating divergent views and acting a counterbalance to scrutinize an incumbent government as it implements its duties to its citizens [27], [28]. However, in recent years this has led to political extremism and widening ideological differences on fundamental issues. To understand this, researchers have used Twitter extensively to mine public views on various social and foreign policies in many countries. Although, only a small body of literature has focused on the opinion of different political groups, for example, in relation to climate change [29]–[33], solar energy [34], immigration [35], [36], COVID-19 vaccines [37], and Russian election interference [38].

Polarization on these topics can be more prevalent when leading up to a major election, as parties aim to differentiate themselves to voters [39], [40]. This can lead to an increase in political intolerance across competing groups. Indeed, propaganda can become political weapons against opponents [41], and negativity on social media can be amplified and spread faster when compared with positive content. Twitter and other social media sites provide a novel and readily available data source for researchers to study political alignments and the evolution of various qualitative narratives and discussions [42]. For example, in the US, Twitter data analysis illustrates the hollowing out of the political center between the Left and the Right [43], [44]. This is also observed in other democracies. For instance, using data for Indian politicians, research has investigated how political representatives differ and/or agree on national issues [45]. Moreover, network analysis applied to Twitter data for politicians, i.e., retweets and responses, indicates a similar pattern to existing political grouping, reinforcing existing political divisions. Similar behavior is also notable in European countries such as Sweden [46].

Although the Twitter platform allows users to express themselves on important issues, social media platforms are known to reinforce niche (often very extreme) viewpoints across the political spectrum, on the left and right [14]. This occurs because Twitter users tend to align with their *a priori* political groups [19], with the reinforcement of existing viewpoints due to an 'echo chamber' effect where alternative views are actively or unconsciously excluded (either by human decision, automated algorithmic preference, or both). Indeed, public declaration of political stance can also be leveraged by political rivals and used in future political campaigns.

## 2.2   Review of applied NLP studies utilizing Twitter data

There are a wide variety of application areas using NLP techniques, particularly by using the Twitter API as a large comprehensive source of linguistic data. Here we summarize a set of relevant papers.

In recent years there has been a substantial assessment of Twitter narratives in relation to the Covid-19 pandemic, making this topic one of the most studied phenomena on Twitter [47]–[52]. For example, there has been a deluge of papers focusing on Covid-19 perceptions and a range of other socio-economic factors related to health disparities, racial disparities [48], [49], [53], and patient-tweeted information regarding drugs or vaccines [37], [49], [54]. There has also been considerable focus on network effects regarding these key topics, such as investigating the influence of leaders in disseminating information and the impact of moral boosting statements during arduous periods of the pandemic [55]. Indeed, appraisal of the emotions which leaders have depicted in Tweets has been examined, focusing on health, news, politics, and other public conversations [56]. Also, scholars have focused on mental health and substance abuse-related incidences using Twitter data due to covid lockdowns and subsequent income losses [57].

There has also been substantial use of Twitter data to explore topics associated with the political sphere, which involve sentiment analysis before or after an election [58], [59]. For example, in Indonesia and Spain, scholars used pre-election Twitter narratives to predict the outcome of a presidential election [58] and the Catalan independence referendum [60], respectively. However, some have raised concerns about the reliability and trustworthiness of Twitter as an information source due to the ability of these narratives to be negatively affected by misinformation and the spread of 'fake news' (often by adversaries) [42], [61]. Online social media sites have been used as an influence engine to manipulate political outcomes and spread misinformation and misconceptions [62]. In recent years, machine learning approaches have been developed for identifying the spread of hate speech in Twitter spaces, including examining how these narratives evolve based on user retweet patterns [42].

There have also been a range of studies that examine sentiment regarding environmental topics, including biodiversity and climate change [29]–[33], [63]. Climate change is a polarized subject among competing political groups, and the Twitter platform allows researchers to study how leaders from each group differ on climate change policies. For example, n-gram analysis has been used to explore the sentiment of words associated with either 'global warming' or 'climate change' [33]. Indeed, this research attempted to answer if the public is aware of the difference between these two terms, the political association of each terminology, and the point of convergence and divergence between the two discussions. Similar assessments also focused on public awareness of biodiversity [63]. Other contentious social issues studied using Twitter data include abortion, LGBTQ, gender issues, and gender-based violence [64]–[68].

The Twitter micro-blogging feature has been utilized effectively as a distributed sensor system to detect natural disasters, such as earthquakes, and their intensities [69]–[72]. Moreover, the collective market mood can also be captured from Tweets. This approach is an established way for business and finance analysts to explore sentiment regarding investment products and thus be able to conclude how the pricing of these options may respond in the future [73], [74]. Twitter data has also been used to assess customer sentiment towards their fixed and mobile broadband providers [75], [76], with negative sentiment often reflecting the known coverage and capacity issues that exist in broadband infrastructure [77]–[79].

Finally, it is worth acknowledging some of the limitations of this type of data in advance. For example, Twitter has traditionally restricted the length of shared texts [80]. Indeed, the short texts force users to express themselves creatively, albeit at times in less logical ways (for example, using vernacular or slang language). Detecting sarcasm is also a challenging known problem, as it may mean estimated sentiment scores are incorrectly allocated [81].

Furthermore, linguistic diversity, language dynamicity, and rapid switching of public dialogues make it difficult for NLP models to adapt unless retrained with a new set of data [80]. Another challenge is analyzing large quantities of unstructured Tweets, which might be difficult to process without using more sophisticated big data tools [82], [83]. Lastly, Twitter algorithms have been pointed out as biased against some political groups and can amplify or suppress Tweet visibility [84].

*Table 1 A review of studies utilizing Twitter data.*

| Author | Year | Topic | Data Type and Date |
|---|---|---|---|
| Bartelt and Elizabeth [66] | 2020 | LGBTQ, Abortion | Public Twitter, Date Unknown |
| Behl et al. [69] | 2021 | Detection of Natural Disasters | Public Twitter, Nepal, 2015 and Italy, 2016 Earthquake, and Covid datasets |
| Budiharto et al. [58] | 2018 | Election Politics | Public Twitter, 2018-2019 |
| Castorena et al. [67] | 2021 | Gender-Based Violence | Public Twitter, 2019 |
| Chamberlain et al. [44] | 2021 | Politics | US Legislators Tweets, Date Unknown |
| Chaudhry et al. [59] | 2021 | Election Politics | Public Twitter, 2020 |
| Conover et al. [43] | 2011 | Political Polarization | Public Twitter, 2010 |
| Criss et al. [49] | 2021 | Covid Vaccines, Ethnicity, Race | Public Twitter, 2020-2021 |
| de Rosa et al. [35] | 2021 | Immigration | Public Twitter, Date Unknown |
| Evkoski et al. [42] | 2021 | Hate speech | Public Twitter, 2018-2020 |
| Falkenberg et al. [31] | 2022 | Climate Change | Public Twitter, 2014-2021 |
| Fitri et al. [65] | 2019 | LGBTQ | Public Twitter, 2019 |
| Garcia et al. [47] | 2021 | Covid | Public Twitter, 2020 |
| Goel et al. [56] | 2021 | Covid | Public Twitter, 2020 |
| Gunnarsson and David [46] | 2014 | Political Polarization | Public Twitter, 2012 |
| Hswen et al. [48] | 2021 | Covid, Ethnic Stigmatization | Public Twitter, 2020 |
| Hswen et al. [53] | 2020 | Ethnic Disparities, Health | Public Twitter, 2013-2016 |
| Huszár et al. [84] | 2022 | Political Polarization | Public Twitter, Date Unknown |
| Jang et al. [30] | 2015 | Climate Change | Public Twitter, Date Unknown |
| Jiang et al. [37] | 2021 | Covid Vaccines | Public Twitter, 2020 |
| Kandasamy et al. [85] | 2021 | Covid | IEEE Covid Tweets, 2020 |
| Khatua et al. [64] | 2019 | LGBTQ | Public Twitter, Date Unknown |
| Kim et al. [34] | 2021 | Energy, Solar | Public Twitter, 2020 |
| Klein and Adam [14] | 2019 | Political Polarization | Public Twitter, 2017 |
| Labonte et al. [86] | 2021 | Energy | Public Twitter, 2017-2018 |
| Luther et al. [38] | 2021 | Foreign Politics, Political Polarization | Public Twitter, 2016 and 2020 |
| Machuca et al. [51] | 2021 | Covid | Public Twitter, 2020 |
| Marcec et al. [54] | 2022 | Covid Vaccines | Public Twitter, 2020-2021 |
| McGregor et al. [87] | 2016 | Gender Bias, Elective Politics | US Legislators Tweets, 2014 |
| Naseem et al. [52] | 2021 | Covid | Public Twitter, 2020 |
| Ohtani and Shimon [63] | 2022 | Biodiversity | Public Twitter, 2010-2020 |
| Osmundsen et al. [61] | 2021 | Political Polarization | Public Twitter, 2018-2019 |
| Özerim et al. [36] | 2021 | Immigration | Public Twitter, 2016 |
| Prabhakar Kaila et al. [50] | 2020 | Covid | Public Twitter, 2020 |
| Rajendiran et al. [73] | 2021 | Stock Prediction techniques | Public Twitter, Date Unknown |
| Rogers et al. [88] | 2021 | Political Identity | Public Twitter, 2015-2018 |
| Rufai et al. [55] | 2020 | Political Leadership | Public Twitter, 2020 |
| Ruz et al. [60] | 2020 | Detection of Natural Disasters | Public Twitter, Chile, 2010 and Catalan, 2017 |
| Sakaki et al. [70] | 2010 | Natural Disasters Detection | Public Twitter, 2010 |
| Samanta et al. [57] | 2023 | Mental Health | Public Twitter, 2022 |
| Sanford et al. [29] | 2021 | Climate Change | Public Twitter, 2019 |
| Schöne et al. [41] | 2021 | Hate Speech | Public Twitter, 2014-2016 |
| Shi et al. [33] | 2020 | Climate Change | Public Twitter, 2009-2018 |
| Solovev et al. [68] | 2022 | Hate Speech, gender, and ethnicity | US 117 Congress Tweets |
| Swathi et al. [74] | 2022 | Stock Prediction | Public Twitter, Date Unknown |
| Yu et al. [32] | 2021 | Climate Change | 115th Congress Tweets |

## 2.3 Review of methodological techniques

Before a researcher may analyze the large amount of unstructured text generated by online social networks, it is necessary to cluster this information into their respective document and topic collections [89]. Topic modeling discovers latent thematic information in large documents [90], [91]. Clustering documents into their respective topics is an essential step in the sentiment computation of unknown pieces of text and is applied in a range of Twitter datasets [41], [47], [50], [63], [92]. Sentiment analysis alone can suffice for known document categories. For Twitter, filtering algorithms are applied to extract specific topic-related Tweets. The latent Dirichlet allocation (LDA) is a popular document clustering algorithm that is frequently used for the analysis of textual content in electronic media, including Twitter data [63], [90]–[92]. LDA effectively analyzes large text corpora, but the performance of this approach degrades when analyzing small text documents, such as Tweets.

Also, upon discovering relational semantics within the textual content, sentiment analysis is used to extract constituted emotions. Indeed, sentiment analysis extracts a user's semantic content in verbal or written communications [21], [93]–[95]. In this case, sentiment analysis extracts emotions expressed in textual data such as Tweets and classifies them as neutral, positive, or negative. Further, sentiment analysis entails quantifying expressed opinion, referred to as subjectivity. There are various approaches to text sentiment analysis, including (i) lexicon-based methods, (ii) concept-based methods, (iii) hybrid methods, and (iv) machine learning-based methods [93], [96]–[98]. Lexicon-based sentiment analysis models are commonly used and are built based on relational dictionaries, matching text features and important words with defined sentiment values [54], [58]. In contrast, learning-based models can be more accurate and are increasingly used for the sentiment classification of Tweets. Learning methods are further categorized into supervised and unsupervised methods. Supervised learning employs rule-based, probabilistic, linear, or decision-tree models [97].

These techniques have been widely applied in the literature. For example, using supervised learning techniques, [61] trained linear support vector classification (L-SVC), logistic regression (LR), naive Bayes (NB), and support vector machine (SVM) models on the Kaggle Sentiment140 dataset for Twitter sentiment analysis. Deep learning multi-layered perceptrons such as convolutional neural networks (CNN), long short-term memory (LSTMs), and the gated recurrent unit (GRU), recurrent neural network (RNN) can also be trained to extract emotions in Tweets [99], [100]. The training data is usually pre-annotated. Significantly, the performance of sentiment analysis models and algorithms is tested against benchmark datasets such as pre-annotated Twitter data, International Movie Database (IMDB) reviews, and Wiki Text [100], [101]. Sentiment analysis of Twitter data is evolving as new models are developed, which are usually more effective, accurate, and easily adaptable. These models can be hybrid, combining different sentiment analysis techniques. For example, using universal language model fine-tuning (ULMFiT) and SVM [101] achieves high accuracy on WikiText-103 and Twitter US airline data. This is true for multilingual language methods, e.g., the cross-lingual language model for Twitter (XLM-T) model [102].

Due to the complexity of Twitter data, extensive research has been conducted into language feature extraction, significantly improving sentiment analysis accuracy. For example, [103] employs a polarity-aware embedding multi-task learning (PEM) model to extract political bias within Twitter political texts. For accurate results and efficient performance of clustering, predictive, or classification machine learning models, it is necessary to preprocess preliminary data. This technique includes the removal of data point redundancies and dimensionality reduction. For example, synonym expansion and negation replacement drastically improve model accuracy [104], [105]. Feature extractions such as n-gram, term frequency-inverse document frequency (TF-IDF), and word embeddings [92] are standard preprocessing algorithms. The n-gram feature processing technique has received much attention, with several studies using it on Twitter data [33], [57], [63], [85], [106].

*Table 2 A survey of sentiment analysis and topic modeling techniques of Twitter data*

| Author | Year | Methods | Data |
|---|---|---|---|
| AlBadani et al. [101] | 2022 | SVM, ULMFiT | WikiText-103, Twitter US Airline |
| Barbieri et al. [102] | 2022 | XLM-T, | Public Twitter, 2020 |
| Bibi et al. [98] | 2022 | Concept-based sentiment analysis, Deep neural network (DNN), NB | Public Twitter, Date Unknown |
| Curiskis et al. [92] | 2020 | LDA, TFIDF, | Twitter and Reddit, Date Unknown |
| Gandhi et al. [100] | 2021 | word2vec, LSTM, CNN | IMDB Movie Reviews and Twitter, Date Unknown |
| Giachanou et al. [95] | 2016 | Hybrid, ML, Lexicon, Graph techniques | Public Twitter, Date Unknown |
| Khan et al. [82] | 2020 | Hadoop, RNN | Public Twitter, Date Unknown |
| Naseem et al. [105] | 2021 | Testing text preprocessing techniques against ML and DNN | Davidson et al., Waseem et al., and Golbeck et al. datasets |
| Nasser et al. [106] | 2021 | N-Gram, LSVM, TFIDF | Public Twitter Covid dataset |
| Rodrigues et al. [83] | 2022 | Flume, Hadoop, Hive, Lexicon, NB classifier | Public Twitter, Date Unknown |
| Sailunaz et al. [94] | 2019 | NB | Public Twitter, Date Unknown |
| Xiao et al. [103] | 2022 | PEM, Named Entity Recognition (NER) | Twitter ELECTION2020, PARLER, TIMME, and Twitter US Legislators 2019-2020 |
| Yadav et al. [104] | 2021 | L-SVC, LR, NB, SVM | Kaggle Sentiment140 dataset |
| Zhang et al. [99] | 2018 | CNN, GRU | Twitter Benchmark datasets |

# 3. Method

This study uses Twitter to investigate the relationship among US political party members on current affairs topics. By collecting Tweets from the US House of Representatives and Senate (comprised of Democrats, Republicans, and Independents), it is then possible to utilize natural language tools to understand underlying sentiments and emotions.

The United States House of Representatives is the lower chamber of Congress. This legislative branch is responsible for formulating policies, which is the blueprint that steers executive governance - drafting bills, debating them, and passing resolutions. Once the lower house passes a bill, and if the Senate ratifies them, it will be sent to the President for consideration.

The method to investigate the research question will progress as follows. Firstly, using the Twitter API (Application Programming Interface) Python client to collect data from Congress member accounts, it is possible to develop a dataset that can be mined for new information.

Secondly, based on the voting record of each member, GovTrack data on the political score of each member is used to infer the political stance of each representative [107]. As the political party affiliation is generally known for each member, Tweets from the same party are aggregated and contrasted with the Tweets from other parties. The method is based on the approach that by analyzing the Tweets from this small group, we can more generally infer the sentiment of the general population subscribing to each party's ideologies.

Finally, we can use techniques to estimate the sentiment of the text data, plot the distributions of these scores based on political affiliation, and carry out formal statistical testing of key metrics. Policy themes are based on international geopolitical issues for which there may be greater bipartisanship (and thus, lower polarization) and domestic social and environmental issues where there may be a higher degree of polarization.

Figure 1 illustrates the method, from data collection to analysis.

*Figure 1 Overview of the method*

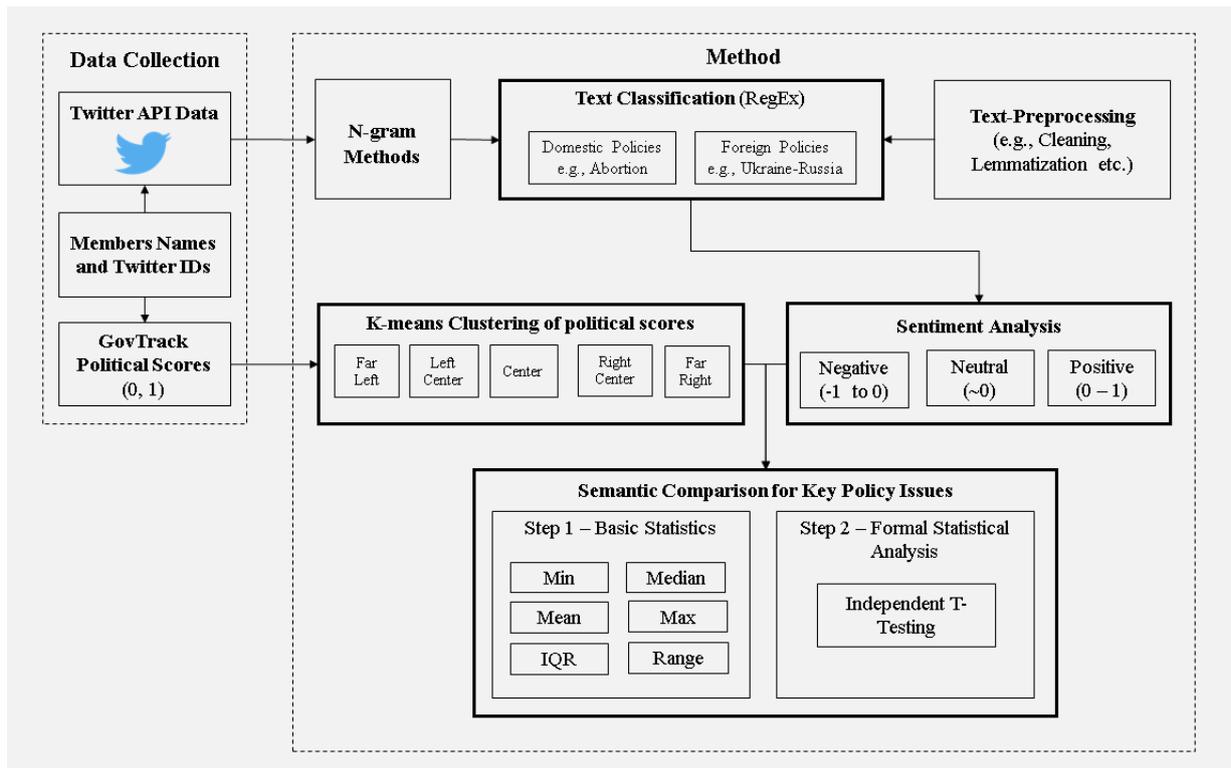

## 3.1 Data

Information about the US congress members was scraped from the web. The data features included the members' names, Twitter handles, party affiliations, and GovTrack scores. Using the Twitter API Python client, *snscrape*, the Tweets from each member were collected between January 2021 and December 2022. From each Tweet, the information extracted was the text (Tweet), date shared, and the Tweet interactivity details (retweets, likes, etc.). The collected Tweets were then saved into a comma-separated values (CSV) file for analysis.

## 3.2 Text Preprocessing

Before sentiment analysis, the fetched Tweet texts are cleaned. This stage involves lowering the text case, expanding the contracted words, i.e., 'don't' to 'do not', removing the hyperlinks embedded in the tweets, removing special characters such as @, #, etc., and removing single characters and white spaces.

## 3.3 Sentiment Analysis

Valence Aware Dictionary for sEntiment Reasoning (VADER) Python package is leveraged to extract the sentiment values from the Tweets [69]. VADER is a rule-based model, and according to its developers, it is effective when used to compute sentiments from social media data. The sentiment output of the model is a normalized value of the sum of the text's negative, neutral, and positive scores and ranges from -1 to +1.

## 3.4 Policy and Political Group Classification

Extracting the underlying topic composition of a large set of textual data is necessary before statistical analysis is undertaken. This section applies an n-gram feature extraction technique to extract the most contiguous terms within the dataset. This is useful as it provides an overview of the currently discussed subjects. The generated sequences of words were one-worded (unigrams), two-worded (bigrams), three-worded (trigrams), and four-worded(fourgrams). The terminologies from n-gram analysis were applied to establish a filtering algorithm that assigned each tweet to a policy category. For example, in the case of extracting tweets related to LGBTQ, the following terms from unigrams and bigrams were used, 'transphobia,' 'homophobia,' 'LGBTQ,' 'trans,' 'biphobia,' and 'sexual identity.'

Finally, applying the K-means clustering algorithm, the quantitative data of each member from GovTrack is used to classify members into a political group. The two main classes, i.e., Democrats and Republicans, were subdivided into five sub-groups, e.g., the Far left, Left Centrist, Centrist, Right Centrist, and Far Right. The GovTrack data range from 0 to 1, with 0 being more politically left and one more politically right.

## 3.5 Statistical analysis

An independent t-test is applied to explore whether the mean sentiment scores between the ideological groups are likely to have occurred randomly or otherwise for a 95% confidence level. No difference is assumed in the mean of the sentiments between the political groups as the null hypothesis for each test.

Using the mean sentiment value for each political group, sets of p-values are obtained using the Python library *statsmodels* to estimate the polarization within the groups. We define polarization as the statistically significant difference of the tested group means within a policy category. The mean sentiments of the political groups on twelve key policies are the basis for comparison and indicate the level of polarization between the groups. The statistical data used [108], and the analysis code and results are available for download.

All method code is available online from the GitHub [109]

# 4. Results

This section reports the descriptive and inferential results examining polarization across political groups on key current issues using sentiment analysis.

## 4.1 Descriptive statistics

Figure 2 illustrates the distribution of elected representatives based on their historical voting record, with the distribution of Democrats in blue and Republicans in red. Regarding basic statistical metrics, the Democrats have a mean political score of 0.28, a modal score of 0.3, a minimum score of 0.00, a maximum score of 0.64, and a range of 0.64. In contrast, the Republicans have a mean political score of 0.71, a modal score of 0.7, a minimum score of 0.44, a maximum score of 1, and a range of 0.56. As the Democrats have a marginally larger range, it suggests their members vote slightly more widely across the political spectrum compared to Republicans. One caveat is that this plot represents the composition of the two houses in 2021, before the midterm elections of 2022.

*Figure 2 Voting record-based political score (lower and upper house elected representatives)*

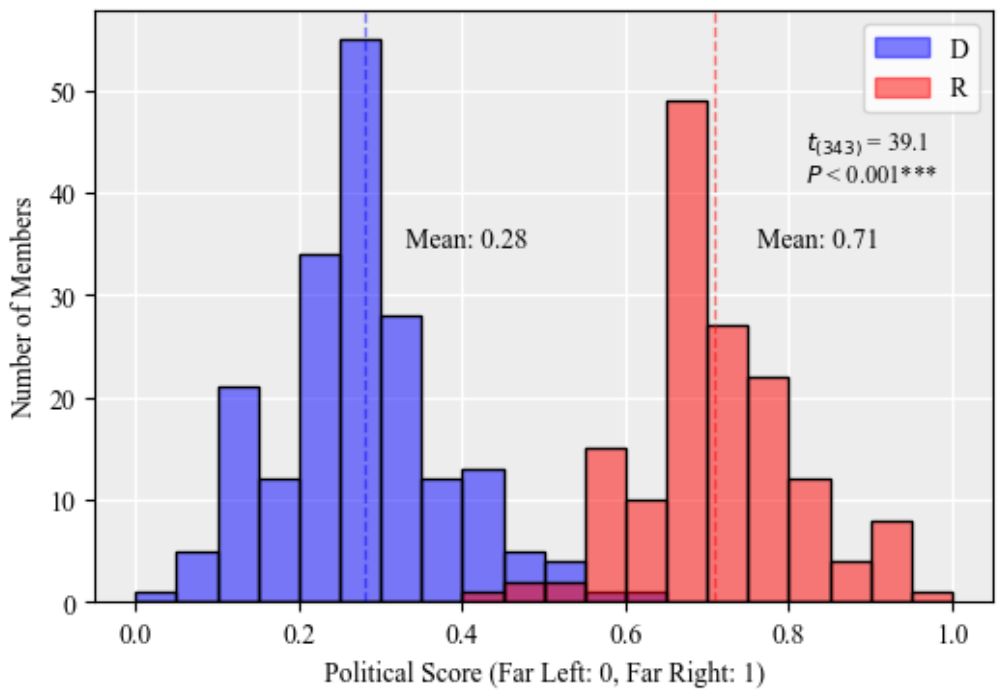

Figure 3 illustrates the frequency of key phrases within the dataset using n-gram methods for single words (unigrams) and phrase sequences consisting of up to a maximum of four (fourgrams). The dataset was collected from Twitter from January 2021 to December 2022.

The most common single word in the dataset is 'biden', which is a similar theme in the two-word phrases, with 'president biden' being the most frequently used. For three-word terms, the most frequent case was 'build back better', which was also top in four-word phrases, based on the phrase 'build back better act'. Compared to the unigram and bigram plots, when allowing for more words per phrase, e.g., with trigrams and fourgrams, we gain much greater insight into the frequency of topics discussed. For example, these include 'bipartisan infrastructure law', 'inflation reduction act', 'woman health protection act', and 'voting rights advancement act', which all provide insight into heavily discussed domestic current affairs topics during the period assessed.

*Figure 3 Frequency composition of common phrases in the dataset using n-gram methods*

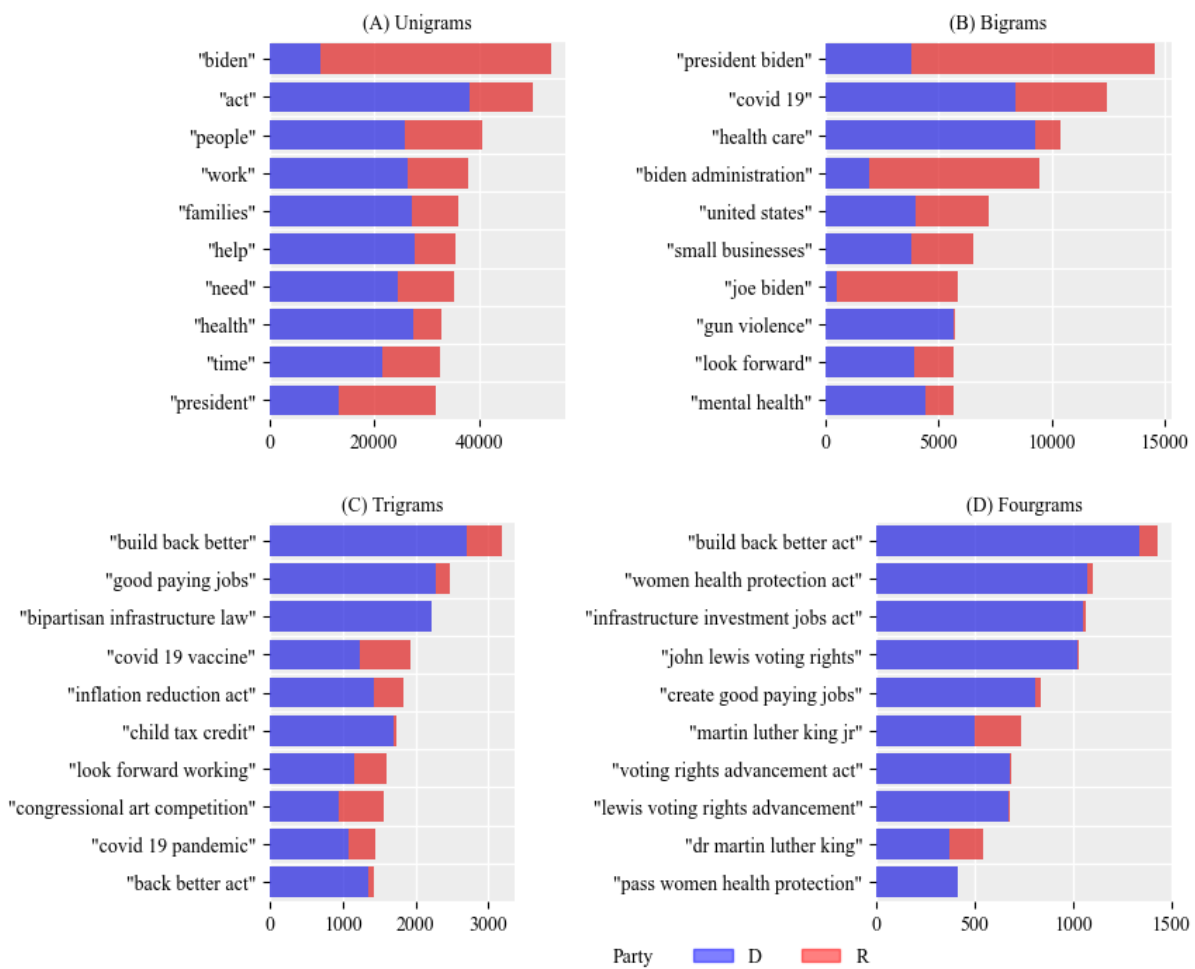

Figure 4 illustrates the estimated sentiment scores for a set of key international geopolitical themes in US politics, including the Chinese Communist Party (CCP), the bipartisan CHIPS and Science Act, Taiwan, and the Ukraine-Russian war. The sentiment is higher for Democrats across the three key areas except for Ukraine-Russia, where sentiments are almost equal. Sentiment outcomes may be driven by the incumbent President's political alignment (thus, comparing these results to results for the previous administration pre-2021 is an important area of further research).

*Figure 4 Key US geopolitical themes and associated sentiment scores*

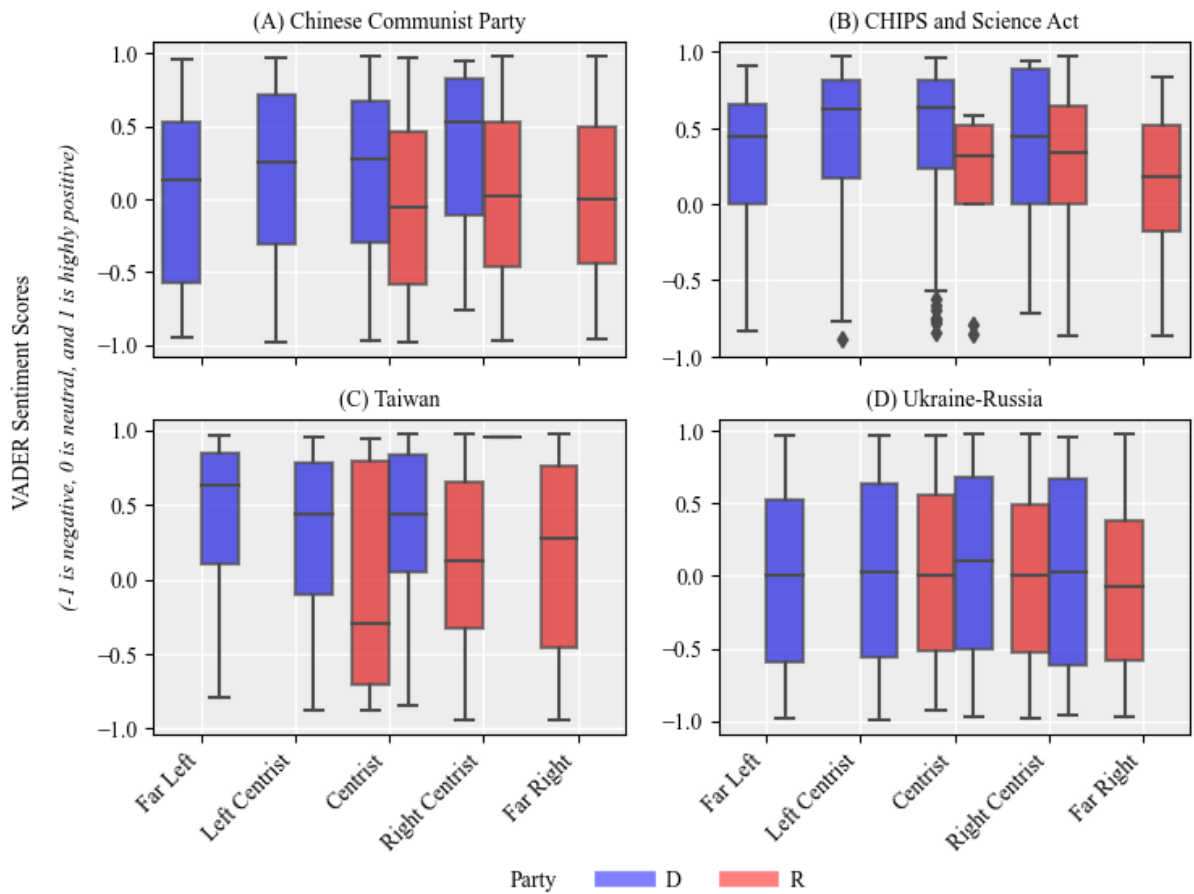

Figure 5 illustrates the sentiment scores across political parties for major US domestic policy themes, including abortion, broadband, climate change, fossil fuels, gun rights, immigration, LGBTQ, and substance abuse. On all issues except gun control, Democrats express more positive sentiment than Republicans.

*Figure 5 Key US domestic policy themes and associated sentiment scores*

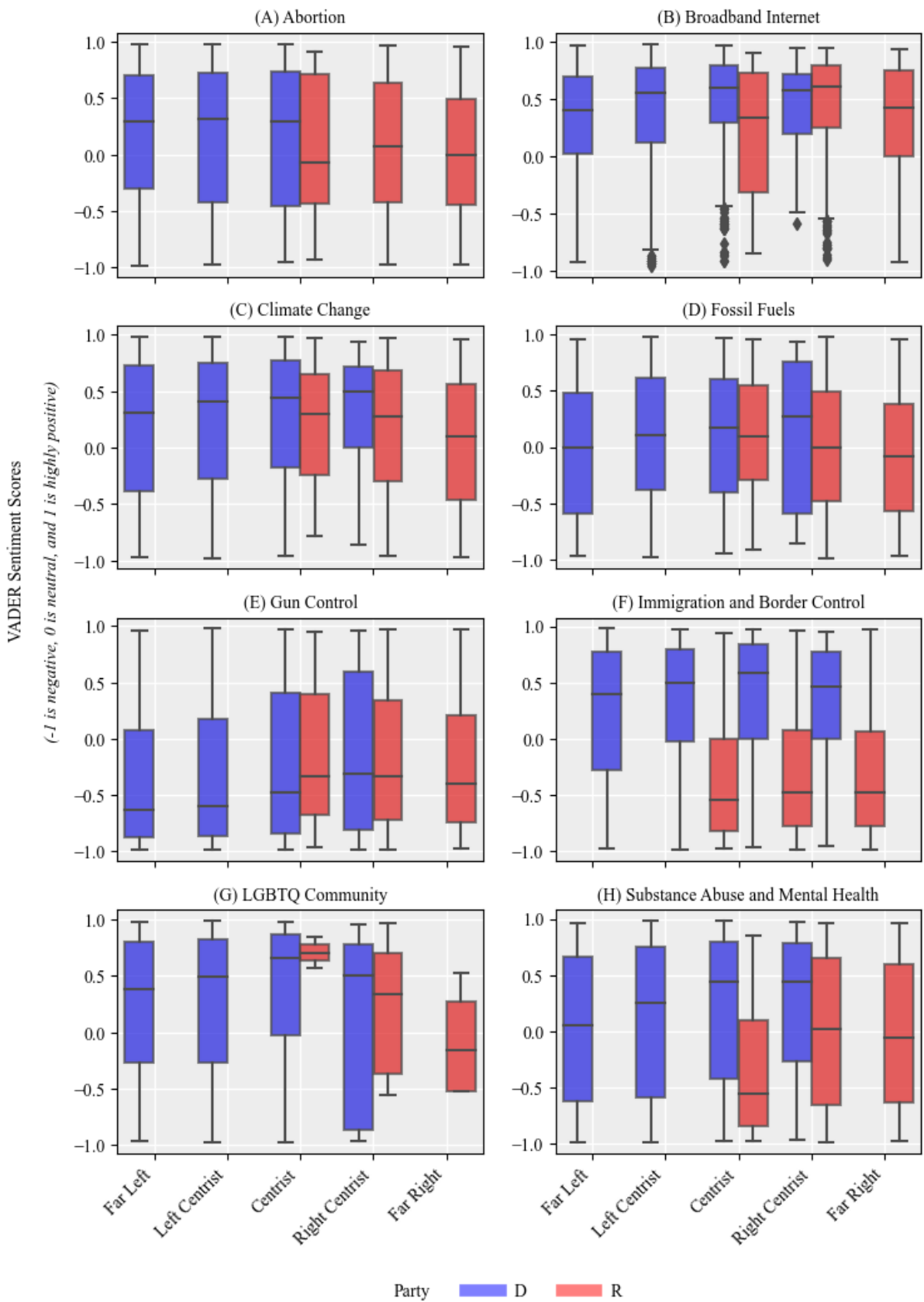

## 4.2 Inferential statistics - Polarization Between Groups

Now the results are reported for the formal testing of polarization on key topics between political groups. Table 3 summarizes the results from the independent t-test of the sentiments across political groupings. The results show that abortion, immigration and border control, gun control, fossil fuels, and Ukraine-Russia discourses have substantive polarization between the political groups tested. The least divisive topics included Taiwan and LGBTQ issues, with no recorded statistically significant difference in sentiment between the political groupings.

When comparing the findings for the sentiment on the topic of abortion, the groups with a significant difference in mean sentiment include Left Centrist and Right Centrist (0.183 vs. 0.095; $p < .001$), the Far Left and the Far Right (0.191 vs. 0.037; $p < .001$), the Far Left and the Right Centrist (0.191 vs. 0.095; $p < .001$), and the Far Right and the Left Centrist (0.037 vs. 0.183; $p < .001$). This demonstrates strong polarization on this issue between opposing groups across the political spectrum. Variance across political groups may be explained by right-leaning elected members generally taking a pro-life stance, compared to left-leaning counterparts generally being more pro-abortion.

Immigration policy and the approach to managing US borders is another highly debated topic among elected representatives. The sentiment comparison of the groups shows a strong significant difference in mean sentiments between the Far Left and the Right Centrist (0.237 vs. -0.288; $p < .001$), and the Far Left and the Far Right (0.237 vs. -0.308; $p < .001$). The difference is also noted with the Left Centrist versus the Right Centrist (0.323 vs. -0.288; $p < .001$) and the Far Right against the Left Centrist (-0.308 vs. 0.323; $p < .001$). Although a difference is realized when the Far Left is contrasted with the Left Centrist, there are very similar positive mean values (0.237 vs. 0.323; $p < .001$). Polarization could be attributed to political ideology whereby left-leaning representatives may be more lenient on how the government should handle illegal immigration. In contrast, right-leaning representatives may take a less tolerant stance, instead advocating for more robust measures to deter future illegal immigration into the US, especially in the US southern border states.

*Table 3 Independent T-test Results*

| Policy | Statistical Metric | Far Left - Far Right | Far Right - Left Centrist | Right Centrist - Left Centrist | Right Centrist - Far Left | Far Right - Right Centrist | Left Centrist - Far Left |
|---|---|---|---|---|---|---|---|
| **Abortion** | DF | 1670 | 2427 | 2603 | 1846 | 1164 | 3109 |
| | T | -5.075 | -4.889 | 3.316 | -3.477 | -1.739 | -0.364 |
| | P | < .001*** | < .001*** | < .001*** | < .001*** | 0.082 | 0.716 |
| **Broadband Internet** | DF | 386 | 1513 | 1763 | 636 | 616 | 1533 |
| | T | -0.546 | -3.649 | -0.096 | 2.569 | -3.023 | 3.052 |
| | P | 0.585 | < .001*** | 0.924 | 0.01* | 0.003** | 0.002** |
| **Chinese Communist Party** | DF | 1865 | 2088 | 2930 | 2707 | 4298 | 497 |
| | T | 0.509 | -2.933 | 2.847 | 0.581 | -0.223 | 1.951 |
| | P | 0.611 | 0.003** | 0.004** | 0.561 | 0.824 | 0.052 |
| **CHIPS and Science Act** | DF | 83 | 323 | 372 | 132 | 115 | 340 |
| | T | -2.237 | -4.771 | 3.284 | -0.396 | -1.965 | 2.248 |
| | P | 0.028* | < .001*** | 0.001** | 0.693 | 0.052 | 0.025* |
| **Climate Change** | DF | 2727 | 5973 | 6490 | 3244 | 1247 | 7970 |
| | T | -2.674 | -4.475 | 0.439 | 1.826 | -3.757 | 3.572 |
| | P | 0.008** | < .001*** | 0.66 | 0.068 | < .001*** | < .001*** |
| **Fossil Fuels** | DF | 1527 | 2147 | 3087 | 2467 | 2514 | 2100 |
| | T | -3.269 | -7.929 | 4.42 | 0.539 | -4.511 | 3.975 |
| | P | 0.001** | < .001*** | < .001*** | 0.59 | < .001*** | < .001*** |
| **Gun Control** | DF | 3144 | 6979 | 7554 | 3719 | 2359 | 8339 |
| | T | 6.086 | 4.982 | -10.601 | 11.083 | -3.143 | 2.229 |
| | P | < .001*** | < .001*** | < .001*** | < .001*** | 0.002** | 0.026* |
| **Immigration and Border Control** | DF | 5781 | 6782 | 9601 | 8600 | 11539 | 3843 |
| | T | -32.15 | -45.375 | 46.446 | -32.002 | -1.904 | 4.451 |
| | P | < .001*** | < .001*** | < .001*** | < .001*** | 0.057 | < .001*** |
| **LGBTQ Community** | DF | 657 | 1599 | 1625 | 683 | 32 | 2250 |
| | T | -0.841 | -0.895 | 1.271 | -1.095 | -0.338 | 0.66 |
| | P | 0.401 | 0.371 | 0.204 | 0.274 | 0.737 | 0.509 |
| **Substance Abuse and Mental Health** | DF | 1228 | 2537 | 2965 | 1656 | 1236 | 2957 |
| | T | -2.538 | -4.505 | 2.964 | -0.543 | -2.077 | 2.33 |
| | P | 0.011* | < .001*** | 0.003** | 0.587 | 0.038* | 0.02* |
| **Taiwan** | DF | 246 | 278 | 261 | 229 | 435 | 72 |
| | T | -1.265 | -0.987 | 1.453 | -1.58 | 0.723 | -0.588 |
| | P | 0.207 | 0.324 | 0.147 | 0.115 | 0.47 | 0.558 |
| **Ukraine-Russia** | DF | 2622 | 4577 | 6086 | 4131 | 4919 | 3789 |
| | T | -2.296 | -6.627 | 2.813 | 1.094 | -4.609 | 2.841 |
| | P | 0.022* | < .001*** | 0.005** | 0.274 | < .001*** | 0.005** |

Gun control evokes tense political discussions between political groups, as supported by the quantitative metrics derived here. When comparing sentiments across groups, the results find that there is a strong difference in sentiments between the Far Left and Right Centrist (-0.386 vs. -0.164; p < .001), the Far Left and the Far Right (-0.386 vs. -0.244; p < .001), the Left

Centrist and the Right Centrist (-0.353 vs. -0.164; p < .001), and the Far Right and the Left Centrist (-0.244 vs. -0.353; p < .001). Moreover, on gun control, the other two groups aligned within the same political segments, the Far Left and Left Centrist (-0.386 vs. -0.353; p = 0.026) and the Far Right versus the Right Centrist (-0.244 vs. -0.164; p = 0.002) exhibit a considerable difference in sentiment for this dataset. Polarization is associated with ideological differences in the regulation of gun ownership, whereby left-leaning representatives may believe in adopting stringent measures that control the ownership of guns, especially assault rifles. In contrast, right-leaning representatives may believe that such controls infringe fundamental constitutional rights should greater government regulation of gun ownership be introduced.

Russia is an adversary nation of the United States, and the 2022 invasion of Ukraine has furthered geopolitical tensions. For example, there have been many debates on national security and economic globalization processes in the United States and across all industrialized democratic nations. The results obtained by analyzing Ukraine-Russian narratives suggest mixed sentiment across the political groups, as shown by the quantitative metrics in Table 3. For example, a significant difference in sentiment mean is found for the Far Left against the Far Right (-0.024 vs. -0.078; p = 0.022), and the Far Left and the Left Centrists (-0.024 vs. 0.043; p = 0.005). Moreover, a similar outcome is obtained when comparing the Left Centrist and Right Centrists (0.043 vs. 0.0; p = 0.005). The Far Right against the Right Centrist (-0.078 vs. 0.0; p < .001), and the Far Right and Left Centrist (-0.078 vs. 0.043; p < .001) show a considerable variation in the sample means, representing strong differences in US policy sentiment on this issue, reflecting opinion differences in military aid contributions. For example, right-leaning representatives may be concerned with the escalation of the war. In contrast, groups on the Left may be more supportive of equipping Ukraine with modern weapons (however, attitudes on this topic are highly heterogenous within each party). Although initially funding Ukraine and supplying it with advanced artillery attracted bipartisanship support in 2022, there are questions from right-leaning representatives regarding corruption and the misuse of funds within the Ukrainian government.

Moreover, another area of heightening geopolitics regards China and the Chinese Communist Party (CCP). Indeed, the CCP is perceived as one of the main political, security, and economic risk posed to the United States, now and over the next century [110], [111]. Based on the data analyzed here, the sentiments are fairly equal between the Left and the Right groups, suggesting little polarization between US, political groupings in this policy category. Indeed, the quantitative data demonstrates that the mean of the Far Left group is lower when compared to those of the Right Centrist (0.029 vs. 0.058; p = 0.561), the Left Centrist (0.029 vs. 0.147; p = 0.052), and the Far Right (0.029 vs. 0.054; p = 0.611). However, there is not a statistically significant difference when explored using independent t-tests of these groups, potentially indicating shared consensus on this issue. This is also true for the Far Right and the Right Centrist (0.054 vs. 0.058; p = 0.824), which exhibit similar means. On the other hand, the Far Right and the Left Centrist (0.054 vs. 0.147; p = 0.003), and the Left Centrist and Right Centrist (0.147 vs. 0.058; p = 0.004) indicate statistically significant differences at the 95% confidence level.

Linked to China is the issue of Taiwanese independence and the current US dependence on microchip production in Taiwan by the Taiwanese Semiconductor Manufacturing Company (TSMC). Only two global foundries have mastered sub-10 nanometer chip production (TSMC and Samsung) [112], [113], introducing a strategic vulnerability for countries that rely on state-of-the-art microchips for economic activities and national security. In recent years, US legislators have pushed toward investment in developing semiconductor foundries and advanced scientific research within the continental United States. For example, the Creating Helpful Incentives to Produce Semiconductors (CHIPS) and Science Act are viewed to attract bipartisan support. The statistical models employed show a significant difference between the Far Right and Left Centrist (0.115 vs. 0.491; p < .001), and the Left Centrist and Right Centrist (0.491 vs. 0.309; p = 0.001). Indeed, a considerable significant difference in the mean is noted between the Far Left and the Far Right (0.343 vs. 0.115; p = 0.028), and between the Far Left and the Left Centrist (0.343 vs. 0.491; p = 0.025). No difference in the mean is observed between the Far Right and Right Centrists (0.115 vs. 0.309; p = 0.052) or the Far Left and the Right Centrists (0.343 vs. 0.309; p = 0.693). Regardless of the inter-political group variation, the average mean of the sentiments is positive overall, suggesting this sentiment supports the bipartisanship that took place in the production of the legislation. Separately, issues pertaining to Taiwanese independence were tested, and results show little political polarization on this key issue. For example, when considering the Far Left group against the Right Centrist (0.41 vs. 0.2; p = 0.115), the Far Left and the Left Centrist (0.41 vs. 0.327; p = 0.558), and the Far Left and the Far Right (0.41 vs. 0.24; p = 0.207) no significant differences were found. This was also true for all other groups, suggesting a strong degree of bipartisanship in Taiwanese independence.

In the literature, right-leaning representatives have had reservations in support of climate change combat efforts compared to the stance of left-leaning political groups. By comparing the mean sentiments of these groups, we find that there is a significant difference between the Far Right and the Left Centrist (0.093 vs. 0.234; $p < .001$), and the Far Right and the Right Centrist (0.093 vs. 0.224; $p < .001$). This is also notable between the Far Left against the Left Centrist (0.182 vs. 0.234; $p < .001$), and the Far Left and the Far Right (0.182 vs. 0.093; $p = 0.008$). Contrary to expectations, there is no statistical significance between the two groups, the Far Left and Right Centrist (0.182 vs. 0.224; $p = 0.068$), and the Left Centrist and the Right Centrist (0.234 vs. 0.224; $p = 0.66$). Disagreements on climate change issues still exist across political groups because there is still a lack of consensus regarding the policy instruments required to combat climate change effects (including on fiscal matters, such as taxation and spending). Furthermore, climate change issues are also strongly related to the use of fossil fuels and the potential shift to cleaner, more sustainable forms of energy. T-test results indicate strong differences in mean values between the left-leaning and right-leaning groups, with the left being more positive than the Right. Indeed, when formally tested, significant differences were found between almost all groups derived here. For example, the Far Left against the Far Right (0.01 vs. -0.083; $p = 0.001$), the Far Left and the Left Centrist groups (0.01 vs. 0.111; $p < .001$), the Left Centrist and the Right Centrist (0.111 vs. 0.023; $p < .001$), as well as the Far right against the Left Centrist groups (-0.083 vs. 0.111; $p < .001$) show a significant difference.

Amidst COVID-19 and rising inflation, there has been growing concern regarding mental health issues, mainly as they may be related to substance abuse. We note a significant difference in mean between the Far Right and the Left Centrist (-0.033 vs. 0.128; $p < .001$) groups on these issues, suggesting strong polarization. Moreover, a substantial difference in means was also found between the Far Right and the Right Centrist (-0.033 vs. 0.048; $p = 0.038$), and the Left Centrist and the Right Centrist (0.128 vs. 0.048; $p = 0.003$). This is also true between the Far Left and the Left Centrist (0.065 vs. 0.128; $p = 0.02$), and the Far Left and the Far Right (0.065 vs. -0.033; $p = 0.011$). There is no difference in mean between the Far Left and the Right Centrist (0.065 vs. 0.048; $p = 0.587$).

Broadband availability in the United States is challenging for many households and businesses, especially those in rural and remote areas. However, there is no consensus on this topic, particularly with regard to how to overcome disparities in broadband infrastructure access. The results on this topic suggest that while there may be a positive sentiment overall across political grouping, there is still a strong degree of polarization. For example, there are statistically significant differences between the Far Left and Left Centrists (0.35 vs. 0.443; p = 0.002), Far Left and Right Centrists (0.35 vs. 0.445; p = 0.01), Far Right and the Left Centrist (0.324 vs. 0.443; p < .001), and the Right Centrists and Far Right (0.445 vs. 0.324; p = 0.003). There is no statistically significant difference between the Far Left and Far Right (0.35 vs. 0.324; p = 0.585), and between the Left Centrist and Right Centrist (0.443 vs. 0.445; p = 0.924).

Finally, the results for LGBTQ issues are examined with no statistically significant differences between the Far Left and the Left Centrist (0.284 vs. 0.303; p = 0.509), the Far Left and the Far Right (0.284 vs. 0.029; p = 0.401), and the Far Left and Right Centrist (0.284 vs. 0.159; p = 0.274). This is also true for the Far Right against the Left Centrist (0.029 vs. 0.303; p = 0.371), and the Far Right and the Right Centrist (0.029 vs. 0.159; p = 0.737). The center groups, the Left Centrist and the Right Centrist (0.303 vs. 0.159; p = 0.204), show no statistical difference in their means.

*Figure 6 Aggregated sentiments of the political groups across the tested policies*

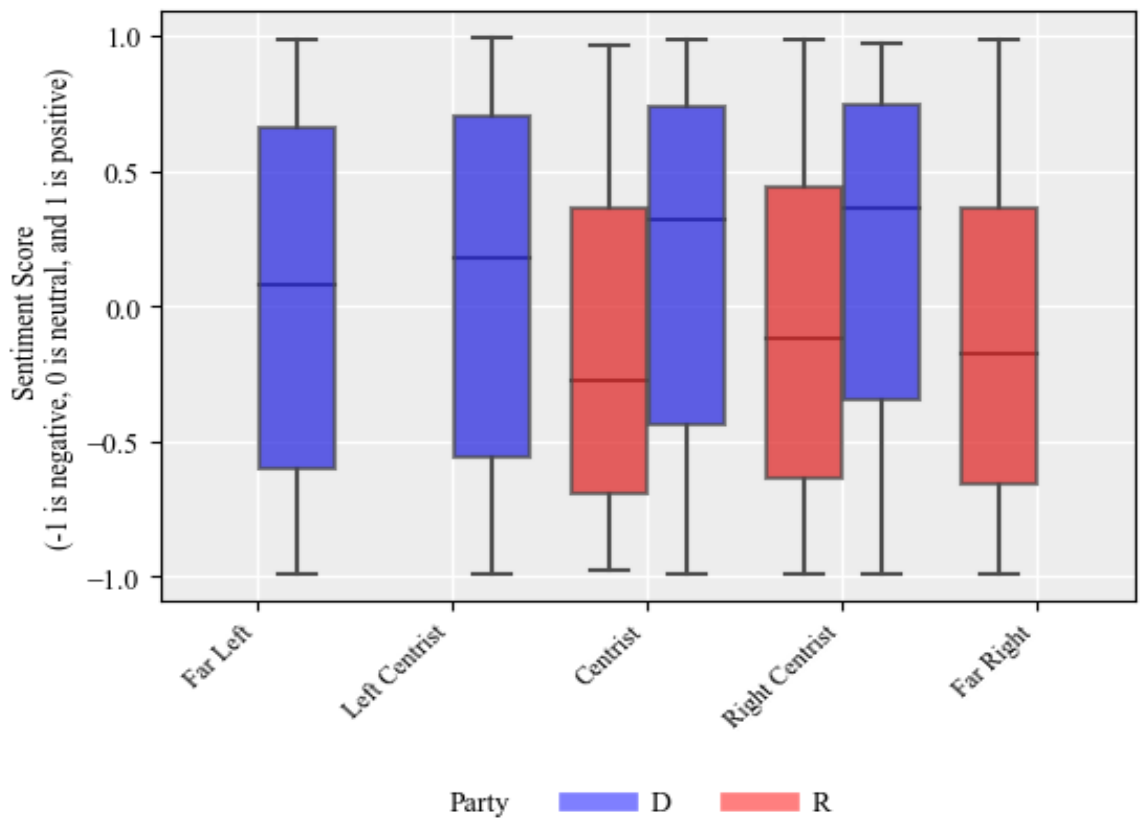

In Figure 7, summary metrics are presented for the degree of polarization found across political groups. The y-axis represents the qualitative policy topic, whereas the x-axis represents the number of statistically significant t-test results for all political groupings where the means were found to differ formally at the 95% confidence level. The shade of each bar component indicates the strength of the significant p-value results found.

The ranking indicates that gun control is the most polarizing policy topic among legislators for the period assessed, with a significant difference across the six political groups, four of which achieved p-values below 0.001. In the second position was immigration which had the largest number of highly significant results by political group, followed by fossil fuels, Ukraine-Russia, and then substance abuse and mental health. In contrast, no significant differences were found in the mean samples of political groups on topics relating to LGBTQ issues and Taiwan, suggesting that these policies are the least polarizing and potentially have the highest degree of bipartisanship.

*Figure 7 Ranking levels of political polarization on key policy issues.*

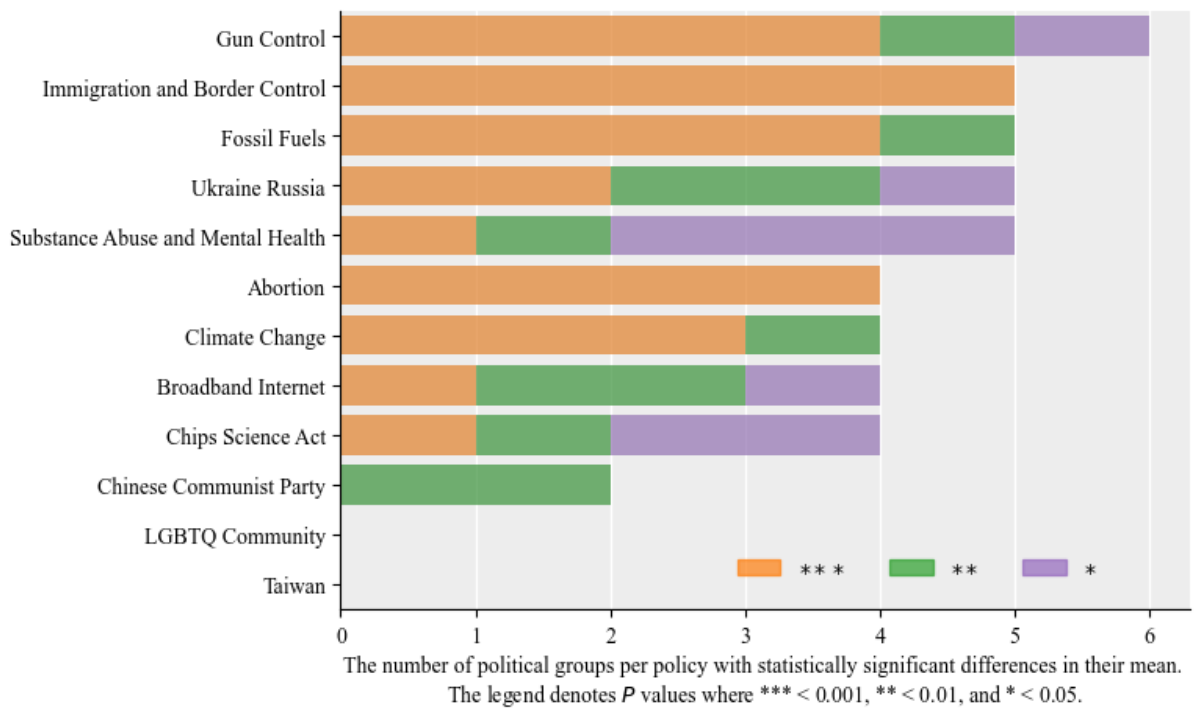

# 5. Discussion

In this discussion, the results will be reviewed with reference to the research question.

*How do the political narratives of different elected representatives compare on key current affairs issues, and to what degree are current political groupings polarized?*

The initial analysis of voting records demonstrated two clear political groups. The bi-modal distribution consisted of Democrat-affiliated representatives with a mean political score of 0.28 and Republican-affiliated representatives having a mean score of 0.71 (where 0 is the most extreme Far Left, and 1 is the most extreme Far Right) ($p<0.001$). Little difference was present in terms of the distributional statistics of these groups, other than Democrats having a marginally larger range, suggesting members vote slightly more widely across the political spectrum compared to Republicans. This voting record distributional information provided important background prior to analyzing the policy topics results.

Subsequently, on the issues assessed sentiment was found to be more positive for Democrats across many of the policy areas examined, as shown in Figure 6, except on topics relating to Ukraine-Russia, where sentiments were practically equal across political groups. Indeed, as on most issues, Democrats expressed a more positive sentiment than Republicans. This suggests there may be a correlation between the incumbent President's political alignment and the sentiment of each group. This is logical as Twitter is used as a public platform to promote or critique policy on key issues, thus making the sentiment outcomes frequently partisan.

The results suggest the most polarizing policy topics across the political groupings explored include gun control, fossil fuels, immigration and border control, Ukraine-Russia, climate change, and abortion. Such findings are commensurate with *a priori* evidence in both the literature and other media outlets. Moreover, the lack of statistically significant differences on key issues such as Taiwanese independence and LGBTQ suggests a high degree of bipartisan support on these topics. Moreover, with only two significant t-test results on the CCP, this suggests a high degree of commonality in the linguistic narratives used by elected representatives for the data analyzed.

# 6.  Conclusions

There has been growing concern that industrialized democratic countries have been experiencing a growing wave of political polarization in recent years. For example, this polarization has been affiliated with greater division and intolerance accelerated by shifts by many parties from the political center ground. It is concerning that strong political divisions could have ramifications for formulating effective policy, potentially leading to more negative societal outcomes.

Therefore, this paper set out to explore the degree of polarization in current political narratives on key policy issues. Using Twitter data between 2021-2022, a sentiment analysis was carried out for elected members of the US House of Representatives and Senate (comprised of Democrats, Republicans, and Independents). After categorizing elected members into political groupings based on past voting records, sentiment values were estimated using Valence Aware Dictionary and sEntiment Reasoner (VADER) for Tweets that contain key policy-related terms. Finally, the dataset developed was explored by formally testing to examine statistical differences in group mean values.

Out of the twelve policy topics explored here, only one topic had statistically significant polarization across all political groups (gun control). However, eight other topics were polarized across either five political groups (immigration and border control, fossil fuels, Ukraine-Russia, and substance abuse and mental health), or four political groups (abortion, climate change, broadband infrastructure, and the CHIPS and Science Act), suggesting these were also highly partisan issues. The least polarized policy topics included Taiwan, LGBTQ, and the CCP.

There are limitations to the work which are important to discuss. For example, the study period between 2021 and 2022 takes place when Democrat-affiliated elected representatives held control of the House of Representatives and Senate and the Presidency. Moreover, prior to the 2020 election, there was a splintering in the usage of social media platforms along party lines, such as Twitter. In the most high-profile case, the incumbent President was banned from the platform, which caused other politically affiliated followers to be less active. Therefore, this could have affected participation across the political spectrum. To overcome these limitations, future research needs to be undertaken which examines the temporal aspects of political

sentiment, by analyzing data from previous periods (e.g., 2014-2016, 2016-2018, 2018-2020, 2020-2022). This would help (i) clarify the relationship between sentiment and control of the Presidency or the House of Representatives and Senate and (ii) provide insight into whether political polarization is increasing or decreasing over time.

**Acknowledgments**

The authors gratefully acknowledge Benjamin Klutsey, Daniel Rothschild, and Martha Anderson and the funding support of the Mercatus Pluralism and Civil Exchange program.

# **Appendix**

*Table A1. Descriptive Statistics*

| Policy | Metric | Far Left | Far Right | Left Centrist | Right Centrist |
|---|---|---|---|---|---|
| Abortion | Count | 1177 | 495 | 1934 | 671 |
| | Mean | 0.191 | 0.037 | 0.183 | 0.095 |
| | Median | 0.318 | 0 | 0.361 | 0.094 |
| | STD | 0.577 | 0.541 | 0.606 | 0.569 |
| Broadband Internet | Count | 204 | 184 | 1331 | 434 |
| | Mean | 0.35 | 0.324 | 0.443 | 0.445 |
| | Median | 0.402 | 0.44 | 0.557 | 0.599 |
| | STD | 0.431 | 0.484 | 0.403 | 0.442 |
| Chinese Communist Party | Count | 138 | 1729 | 361 | 2571 |
| | Mean | 0.029 | 0.054 | 0.147 | 0.058 |
| | Median | 0.115 | 0.024 | 0.25 | 0.052 |
| | STD | 0.601 | 0.536 | 0.601 | 0.551 |
| CHIPS and Science Act | Count | 51 | 34 | 291 | 83 |
| | Mean | 0.343 | 0.115 | 0.491 | 0.309 |
| | Median | 0.45 | 0.115 | 0.637 | 0.382 |
| | STD | 0.453 | 0.467 | 0.43 | 0.491 |
| Climate Change | Count | 2363 | 366 | 5609 | 883 |
| | Mean | 0.182 | 0.093 | 0.234 | 0.224 |
| | Median | 0.318 | 0.166 | 0.402 | 0.351 |
| | STD | 0.595 | 0.577 | 0.582 | 0.555 |
| Fossil Fuels | Count | 741 | 788 | 1361 | 1728 |
| | Mean | 0.01 | -0.083 | 0.111 | 0.023 |
| | Median | 0.026 | -0.077 | 0.128 | 0 |
| | STD | 0.569 | 0.536 | 0.55 | 0.548 |
| Gun Control | Count | 2253 | 893 | 6088 | 1468 |
| | Mean | -0.386 | -0.244 | -0.353 | -0.164 |
| | Median | -0.649 | -0.402 | -0.623 | -0.311 |
| | STD | 0.594 | 0.587 | 0.614 | 0.602 |
| Immigration and Border Control | Count | 1422 | 4361 | 2423 | 7180 |
| | Mean | 0.237 | -0.308 | 0.323 | -0.288 |
| | Median | 0.402 | -0.48 | 0.511 | -0.477 |
| | STD | 0.597 | 0.541 | 0.562 | 0.558 |
| LGBTQ Community | Count | 655 | 4 | 1597 | 30 |
| | Mean | 0.284 | 0.029 | 0.303 | 0.159 |
| | Median | 0.468 | 0.058 | 0.542 | 0.384 |
| | STD | 0.607 | 0.466 | 0.613 | 0.745 |
| Substance Abuse and Mental Health | Count | 825 | 405 | 2134 | 833 |
| | Mean | 0.065 | -0.033 | 0.128 | 0.048 |
| | Median | 0.103 | -0.048 | 0.291 | 0.103 |
| | STD | 0.636 | 0.643 | 0.661 | 0.646 |
| Taiwan | Count | 21 | 227 | 53 | 210 |
| | Mean | 0.41 | 0.24 | 0.327 | 0.2 |
| | Median | 0.681 | 0.382 | 0.44 | 0.307 |
| | STD | 0.587 | 0.588 | 0.528 | 0.58 |
| Ukraine-Russia | Count | 918 | 1706 | 2873 | 3215 |
| | Mean | -0.024 | -0.078 | 0.043 | 0 |
| | Median | 0 | -0.052 | 0.026 | 0 |
| | STD | 0.612 | 0.553 | 0.617 | 0.567 |